# On the Existence of Quantum Wave Function and Quantum Interference Effects in Mental States: An Experimental Confirmation during Perception and Cognition in Humans
# On the Possibility That We Think in a Quantum Mechanical Manner: An Experimental Verification of Existing Quantum Interference Effects In Cognitive Anomaly of Conjunction Fallacy

Elio Conte[1,2], Andrei Yuri Khrennikov[3], Orlando Todarello[4], Antonio Federici[1], Roberta De Robertis[4], Joseph P. Zbilut [5]

1. Department of Pharmacology and Human Physiology – TIRES – Center for Innovative Technologies for Signal Detection and Processing, University of Bari- Italy;
2. School of Advanced International Studies for Applied Theoretical and Non Linear Methodologies of Physics, Bari, Italy;
3. International Center for Mathematical Modeling in Physics and Cognitive Sciences, MSI, University of Växjö, S-35195, Sweden;
4. Department of Neurological and Psychiatric Sciences, University of Bari, Italy;
5. Department of Molecular Biophysics and Physiology, Rush University Medical Center, 1653 W, Congress, Chicago, IL 60612, USA.

   Correspondence to: Elio Conte   (email: elio.conte@fastwebnet.it) .

**Abstract:** We introduce the quantum theoretical formulation to determine  a posteriori, if existing, the quantum wave functions and to estimate the quantum interference effects of mental states. Such quantum features are actually found in the case of an experiment involving the perception and the  cognition in humans. Also some specific psychological variables are introduced and it is obtained that they characterize in a stringent manner the quantum behaviour of mind during such performed experiment.

**1.Introduction**
Mental operations  have  a content plus the awareness of such content. Consciousness is a system which observes itself , and evaluates itself being aware at the same time of doing so. Statements may be indicated by a, b, c, … , .They  are self-referential or auto referential. Content statements of our experience may be expressed by  x , y, z, … . According to A.G. Kromov [1],
$a = F(a, x)$
is the most simple definition of a single auto referential statement.
As an example, consider
 x = the sun shines in the sky;
a=I am aware of this.
 Human experience unceasingly involves  our  perception-cognition system .  Mental and experiential functions such as "knowing" and "feeling"  are involved with sensory inputs, intentions, thoughts and beliefs. A continuous interface holds between mind/consciousness and brain.
Neuroscience and neuro-psychology have reached  high levels of  knowledge in this field by the extended utilization of electrophysiological and of functional brain imaging technology. However, neuroscience finds  it hard to identify the crucial link existing between empirical studies that are currently described in psychological terms and the data that arise instead described in neurophysiological terms. It is assumed that the measurable properties of the brain through functional imaging technology should be in itself sufficient to achieve an adequate explanation of the psychologically described phenomenology that occurs during neuropsychological experiments.

Instead, some investigators suggest that intrinsically mental and experiential functions such as "feeling" and "knowing" cannot be described exclusively in terms of material structure, and they require an adequate physics in order to be actually explained. To this purpose they outline the important role that quantum mechanics could carry out. In particular, we outline here the effort of Stapp in several years and still more recently [2], and the prospects for a quantum neurobiology that were outlined already more than a past decade ago [3]. Therefore, it becomes of fundamental and general interest for neuroscience and neuro-psychology to indicate by results of experiments if quantum mechanics has a role in brain dynamics. In the present paper we give a contribution concerning this basic problem while previous results obtained by us on this matter, are given in ref.[4]. First we consider the problem to acquire (a posteriori) a knowledge of quantum wave function starting directly from experimental data, and soon after we show that mental states follow quantum mechanics during perception and cognition of figures having intrinsic ambiguity.

## 2. Quantum Theoretic Approach

Quantum mechanics represents the most celebrated theory of science. Started in 1927 by founder fathers as Bohr, Heisenberg, Schrödinger, and Pauli [5], it has revolutionized our understanding of physical reality. It was introduced to describe the behaviour of atomic systems but subsequently its range of validity has turned out to be much wider including in particular some macroscopic phenomena. The conceptual structure and the axiomatic foundations of quantum theory repeatedly suggested from its advent and in the further eighty years of its elaboration that it has a profound link with mental entities and their dynamics. We retain that this feature represents an important connotation of the theory also if it is necessary a correct interpretation of the connection between quantum mechanics and mental properties in the sphere of our reality. One cannot have in mind a quantum physical reduction of mental processes. N. Bohr [6] borrowed the principle of complementarity, which is at the basis of quantum mechanics, from psychology. He was profoundly influenced from reading the "Principles of Psychology" by W. James [7]. However, N. Bohr had not in mind quantum-reductionism of mental entities. Starting with 1930, there was also an important correspondence between W. Pauli and C.G. Jung that culminated in the formulation of a theory of mind-matter synchronization [8]. Also in this case these authors were distant to consider a quantum-reductionism perspective. V. Orlov [9] proposed a quantum logic to describe brain functions but also he did not look for reduction of mental processes to quantum physics. The correct way to frame the problem is not to attempt a quantum reduction of mental processes. We retain that we may arrive to give experimental evidence that cognitive systems are very complex information systems, to which also the laws of quantum systems must be applied. This should represent an important result since it might give a further chance to represent and to understand the principles and rules acting as counter part of human mind.

Let us introduce now the basic framework of our formulation.

The wave function $\psi$ of quantum mechanics represents a mental object. We have the problem of its determination that is to say the manner in which we may acquire a posteriori a knowledge about the fact that a given system is described by a function $\psi$. Starting with 1983 [10] and 2004 [11] we considered this problem for biological and mental exploration. The basic key is that wave function $\psi$ is not an observable in the usual sense of quantum theory and, consequently, it cannot be measured in the usual sense of this word. On the other hand, it may be determined provided that one has an ensemble of similarly prepared systems, each describable by the same function $\psi$. To this purpose [10, see also ref.11] one measures an arbitrary but complete set of observables that we call A on a large number of systems and get a statistical distribution of eigenvalues $A_n$. This approach determines the absolute values $|a_n|$ of the coefficients of the decomposed wave function

$$\psi = \sum_n a_n \psi_{A_n} \quad \text{with} \quad \sum_n |a_n|^2 = 1 \tag{1}$$

In order to determine $\psi$ one must know not only the absolute values of the coefficients but also the phases

$$a_n = |a_n| e^{i\alpha_n} \tag{2}$$

Therefore, we have to repeat the same procedure again by measuring another complete set of observables B on a large number of systems to obtain an independent decomposition

$$\psi = \sum_m b_m \psi_{B_m} \quad \text{with} \quad \sum_m |b_m|^2 = 1 \tag{3}$$

and determine $|b_m|$. Let us admit that we find a number $N$ and $N'$ of coefficients different from zero, respectively. In the case $N > N'$ we introduce the decomposition

$$\psi_{B_m} = \sum_n c_{mn} \psi_{A_n} \tag{4}$$

in the (3), and comparing it with the (1), we have that

$$a_n = \sum_m b_m c_{mn} \tag{5}$$

We have a set of $N$ complex equations, $2N$ real equations for $N + N'$ phases to be determined. If $N' > N$, we have to interchange $A$ with $B$. If they result not soluble, it means that we have not a pure quantum state but a mixture. If instead they result dependent so that the phases cannot be determined uniquely, this means that A and B are not independent of each other and consequently we have to make a new choice for observables.

In conclusion, we have delineated the method to realize a determination of wave functions of mental states starting directly from experimental data. The approach must be clear. It is possible to determine the wave function only post factum and considering an ensemble of similarly prepared subjects. In any case the concept of measurement of observables applied to single systems cannot be confused with the estimation of a wave function. This last is expected to represent the system but remains unknown. In other terms, the function $\psi$ is not observable in itself and thus cannot be measured. However, being subject to probabilistic interpretation, it may be estimated statistically. This is to say that we make only an a posteriori statistical reconstruction of it.

### 3. The Quantum Formulation of Mind Entities.

According to Eccles and Beck [12] the mind is a field of probability. In quantum mechanics, the abstract probability fields causally preside over the advent of events in nature. Mathematically, a quantum state $\psi$ (wave function) is a vector in Hilbert space $H$ over some field. Such a state has a probabilistic content, the vector that represents it, has to be of length one, that is to say $<\psi/\psi>^2 = 1$ (where $<\cdot/\cdot>$ denotes the inner product in $H$). The basic key of intrinsic indetermination in quantum mechanics is the principle of superposition. It states that for any two states $\varphi_1$ and $\varphi_2$ of $H$, there exist at least one other state $\psi$ of $H$ that can be written as a linear combination of the first two:

$$\psi = c_1 \varphi_1 + c_2 \varphi_2 \tag{6}$$

with $|c_1|^2 + |c_2|^2 = 1$ ; $<\varphi_i / \varphi_j> = \delta_{ij}$ ($i = 1,2$ ; $j = 1,2$) \hfill (7)

and $|c_1|^2$ represents probability for state $\varphi_1$ and $|c_2|^2$ is probability for state $\varphi_2$.

Mind states are represented by the quantum superposition principle.

In our quantum model of mind entities (for details see also [4]) we consider that an human subject can potentially have multiple mind representations of a given choice situation, also if actually he can attend to only one representation at any given time. In this quantum mechanical framework we distinguish a potential and an actual or manifest state of consciousness. The state of the potential consciousness will be represented by a vector in Hilbert space. If we indicate for example a bi dimensional case of a decision situation with potential states $|+>$ and $|->$, in relation to a dichotomous observable $A = +,-$, the potential state of consciousness will be given by the superposition

$$\psi = a|+> + b|->. \tag{8}$$

Here, $a$ and $b$ represent probability amplitudes so that $|a|^2$ will give the probability that the state of consciousness, represented by $|+>$, will be finally actualised or manifested during decision. Conversely $|b|^2$ will represent the probability that state $|->$ of consciousness will be actualised or manifested during decision. It will be $|a|^2 + |b|^2 = 1$.

The potential state, given in (8), represents in some manner the condition of doubt or of inner conflict or of intrinsic indetermination of the human subject in relation to the posed question ($A = +,-$).
The amount of doubt of the subject (see also [9]) or his inner conflict or indetermination, in relation to the posed question ($A = +,-$), is given by

$$D = 1 - \left||a|^2 - |b|^2\right| \; ; \; 0 \leq D \leq 1 \tag{9}$$

When an actual or manifest state of consciousness is realized during decision, the (8) is reduced to
$|+>$ with probability $|a|^2$ or to $|->$ with probability $|b|^2$.

As neurophysiological counterpart of the present quantum mechanical model of mind entities, as also previously outlined in [13], we admit that, when a conscious decision or observation happens, the actual event that in correspondence is realized in consciousness, is linked to a particular neural correlate brain state. In this manner, in the (8), $|+>$ and $|->$ represent two possible states having two distinct neural correlates.

For brevity we will not consider here the case of the evolution in time of the state of potential consciousness, see the [4] for details.

Some comments may be now added to the previous formulation. The first is that this quantum model of mind entities must be confirmed experimentally in order to be accepted. It does not exist experimental evidence that states of mind may be represented as vectors in a Hilbert space as well as it does not exist experimental evidence that mind states may be represented by quantum wave functions. We retain that the experimental results obtained in this paper give a satisfactory evidence of this matter. Therefore, the consequences of such result are important. The principle of superposition of states implies that the state space is non- Boolean. Therefore, mind states pertain to a non-Boolean state space. The arising conclusion is that at least some perceptive- cognitive systems have such quantum-like abilities. The brain should result to emulate quantum dynamics at least under some conditions. Such an emulation would allow for a three-valued logic in human cognition: true, false and the superposition of true and false. This could explain the peculiar human ability to hold contradictory notions in mind simultaneously, although usually there is collapse to one state or the other. This ability to see things from "opposite" views might have been valuable in the development of sociability, empathy and even cognitive innovation which seems to depend on seeing things in a radically different way as compared to social or cultural norms. The other important feature relates the nature of mind entities. If the previous model is confirmed, we have to conclude that they, at least under our conditions of experimentation, operate by quantum probabilities and analysing (even unconsciously) probabilities of various alternatives. They work directly with mental wave functions or probabilistic amplitudes.

**4. Theoretical Description of the Performed Experiment.**

In our formulation a decision is asked to a subject on a question ($A = +,-$) selecting it among a set of potential alternatives. According to quantum mechanical results that we have given in the section of the quantum theoretic approach, it becomes of interest to establish the dynamics of such decision mechanism when two or more questions $A, B,...$ are posed to the human subject. As said previously, we associate to every decision situation an observable that we denote by $A, B,...$ that we consider to act on $H$. We may study more that one decision situation, say ($A = +,-$) and ($B = +,-$) to be posed in rapid succession to the subject. We know that a key question in quantum mechanics is whether the corresponding observables are or not commuting operators in $H$, i.e., whether $AB = BA$. From a formal view point we have discussed the case $AB \neq BA$ by the formulas (1-5) of the previous sections. From a cognitive view point we may expect that if $A$ and $B$ commute, a decision $A$ will not affect the subsequent decision on $B$. The situation is completely different in the case in which observables $A$ and $B$ do not commute. We know that in this case the predictions of quantum mechanics differ radically from those of the classical probabilistic model. In this case the quantum probability calculus generates cross-terms also called the interference terms. In our formulation these cross-terms are the signature of an existing intrinsic indetermination, of an intrinsic doubt, of an inner conflict that characterizes the cognitive status of the subject in the sequence $A$ and B of posed questions. Generally speaking, admitting that questions $A$ and $B$ have the same number $n$ of possible choices, considering again the (1-5), we obtain

$$p_B(b_n) = \left(\sum_{j=1}^{n} a_j c_{nj}\right)^2 = \sum_{j=1}^{n} a_j^2 c_{nj}^2 + 2\sum_{j \neq j'} \left[(a_{j'} c_{nj})(a_j c_{nj'})\right] \qquad (10)$$

where the first term in the (10) indicates the classical term of Bayes probability to which it is added the interference term that is the quantum expression of an irreducible intrinsic indetermination during cognition of the subjects under investigation. In brief, if quantum mechanics has a role in the investigated decision process, we have a violation of the classical Bayes formula for conditional probability. Let us examine the case of two dichotomous observables and thus involving the sequence of two decisions, ($A = +,-$) and subsequent (B=+,-). We have the following wave functions

$\psi(A = +) = \sqrt{P(A = +)}\ e^{i\varphi_1}$, $\qquad\qquad \psi(A = -) = \sqrt{P(A = -)}\ e^{i\varphi_2}$,

$\psi(B = +) = \sqrt{P(B = +)}\ e^{i\vartheta_1}, \psi(B = -) = \sqrt{P(B = -)}\ e^{i\vartheta_2}$ \qquad (11)

According to the (1-5) we have that

$\psi(B = +) = \sqrt{P(B = +/A = +)}\ \psi(A = +) - \sqrt{P(B = +/A = -)}\ \psi(A = -)$

$\psi(B = -) = \sqrt{P(B = -/A = +}\ \psi(A = +) + \sqrt{P(B = -/A = -)}\ \psi(A = -)$ \qquad (12)

The square modulus of such probability amplitudes gives

$P(B = +) = P(A = +)P(B = +/A = +) + P(A = -)P(B = +/A = -)$

$- 2\sqrt{P(A = +)P(A = -)P(B = +/A = -)P(B = +/A = +)}\ \cos(\varphi_1 - \varphi_2)$ \qquad (13)

and

$$P(B=+) = P(A=+)P(B=-/A=+) + P(A=-)P(B=-/A=-)$$
$$+ 2\sqrt{P(A=+)P(A=\_)P(B=-/A=+)P(B=-/A=-)}\cos(\varphi_1-\varphi_2) \quad (14)$$

We see that in the (13) and (14), in addition to the classical Bayes formula for conditional probabilities, an interference term appears that acclaims the presence and the role of quantum effects in the sequence B/A investigated at the cognitive level of the subject. The experimental determination of probabilities $P(B=+)$, $P(B=-)$, $P(A=+)$, $P(A=-)$, $P(B=+/A=+)$, $P(B=+/A=-)$ enables us to calculate the interference term

$$\cos\omega = \frac{P(B=+) - P(A=+)P(B=+/A=+) - P(A=-)P(B=+/A=-)}{2\sqrt{p(B=+)p(A=+/B=+)p(B=-)p(A=+/B=-)}} =$$
$$= \frac{\Delta p}{2\sqrt{P(A=+)P(B=+/A=+)P(A=-)P(B=+/A=-)}} \quad (15)$$

and the phase $\omega = \varphi_1 - \varphi_2$ by which the a posteriori determination of the quantum wave function (see the 12) is realized.

Finally, we intend to introduce here some new variables that may be able to characterize the mental condition of a subject in the course of an experiment employing two non commuting dichotomous observables $A$ and $B$.

The first variable was previously given in (9). As previously said, it characterizes the amount of doubt, of inner conflict or of intrinsic indetermination of the subject, and it may be calculated also in the case of a sequence of posed questions $A$ and then $B$. $D_{B/A}$ will characterize in this case the amount of doubt of the subject also in relation to the influence induced on decision on $B$ from previous observation of $A$. In this case we may have quantum interference. In order to quantify such kind of mental effect we may introduce two new variables. They are

$$D' = |P(A=+) - P(A=-)| \quad ; \quad I_{\max} = \left|\sqrt{P(A=+)} + \sqrt{P(A=-)}\right|^2 \quad ;$$
$$I_{\min} = \left|\sqrt{P(A=+)} - \sqrt{P(A=-)}\right|^2 ;$$
$$V = \frac{I_{\max} - I_{\min}}{I_{\max} + I_{\min}} \quad ; \quad D'^2 + V^2 \leq 1. \quad (16)$$

They characterize the psychological condition of quantum interference during the sequence $(B/A)$ of perception-cognition of the subjects in our experiment. They are introduced in analogy with duality relations of physics [14].

$D' \approx 0$ gives maximum quantum interference effect with $V \approx 1$ for intensity of quantum interference. $D' \approx 1$ and $V \approx 0$ express instead low interference effects and low intensity of such psychological effect. We intend to outline the relevance of such novel relations in psychophysics analysis of perception-cognition experiments based on the human elaboration of sequences $A$ and $B$. They are able to quantify quantum effects in a given sequence $A$ and $B$, or also in a relative manner for a group of different sequences ($A$ and $B$), ($A_1$ and $B_1$),……….., ($A_n$ and $B_n$) comparing the different psychological effects that are induced each time by the different sequences.

### 5. The Experiment
Our experiment related the perception-cognition system of human subjects.

Studies on perception indicate that the mental representation of a visually perceived object at any instant is unique even if we may be aware of the possible ambiguity of any given representation. A well known example is the Necker cube [15]. We see the cube in one of the two ways and only one of such representations is apparent at any time. We may be able to see the ambiguity of the design and even we may be able to switch wilfully between representations: we can be aware that multiple representations are possible but we can perceive them only one at time, that is serially. Bistable perception is induced whenever a stimulus can be thought in two different alternatives ways. In our quantum like model of mental states [see also ref.4] we consider that an individual can potentially have multiple representations of a given choice situation, but can attend to only one representation at any given time. In this quantum mechanical framework we distinguish a potential and an actual
or manifest state of consciousness.

Let us consider two ambiguous figures as given in Fig. 1 . A and B are two dichotomous questions which can be asked of people, S, with possible answers "yes (+) or not (-)". We consider A and B to represent two mental quantum like observables of people S under investigation. We split the given ensemble S of humans into two sub ensembles U and V of equal numbers. To ensemble U we pose the question $B$ with probability in answering, given respectively by $P(B=+)$ and $P(B=-)$, and $P(B=+)+P(B=-)=1$. We pose the question $A$ immediately followed by the question $B$ to the ensemble V. We calculate the conditional probabilities $P(B=+/A=+)$ and $P(B=+/A=-)$ and corresponding probabilities for the case $(B=-)$. We reach in this manner a no evadable feature of such experiment. If we obtain $\Delta p \neq 0$ as given in the (15), we certainly are in presence of two no commuting quantum observables ($AB \neq BA$), and we may estimate quantum interference effects and the mental wave function.

We analysed a group of 72 subjects giving geometrical figures (Fig.1) as Test A and Test B, respectively. All the subjects were selected with about equal distribution of females and males, aged between 19 and 22 years. The ambiguity induced by tests of Fig.1 was ascertained for each subject after their answers to question $(A=+,-)$, and $(B=+,-)$. All had normal or corrected-to-normal vision. All they were divided by random selection into two groups (1) and (2). Group (1) was subjected to test $B$ only, while the group (2) was subjected to Test $A$ and soon after (about 800 msec. after choice for test $A$) to test $B$. Each subject was asked to select A=+ or A=- (respectively B=+ or B=-) on the basis of what he was thinking about the figure at the instant of observation. A constant visual angle $V = 2arctg(S/2d) = 0.33\,rad.$ was used with $S$ object's frontal linear size and $d$ distance from the center of the eyes for all the subjects. The figures were placed in front of the eyes of the observer at a distance of 60 cm, and illuminated by a lamp of 60 W located above and behind the observer's head. The experimental room was kept under daylight illumination.

## 6. Results and Conclusions.

We obtained the following results. For group (1) with test B only : P(B=+)=0.6667 ; P(B=-)=0.3333.
For group (2) with test A and soon after test B : P(A=+)=0.5556 ; P(A=-)=0.4444; P(B=+/A=+)=0.6000;P(B=+/A=-)=0.3750;P(B=-/A=+)=0.4000;P(B=-/A=-)=0.6250.
Bayes formula for conditional probabilities gave
P(A=+)P(B=+/A=+)+P(A=-)P(B=+/A=+) = 0.5000;
P(A=+)P(B=-/A=+)+P(A=-)P(B=-/A=-)=0.5000.
Consequently, we had $\Delta p(B=+) = 0.1670$; $\Delta p(B=-) = -0.1667$, $\cos\omega = -0.35363$ for $(B=+)$, $\cos\omega = -0.33549$ for $(B=-)$. A statistical analysis was performed . We had a chi-square value $\chi^2 = 5.7143$ with a satisfactory statistical significance , $\alpha = 0.0168$ (*) , df=1.

The obtained results enable us to confirm that we had quantum interference effects during perception-cognition of figures having intrinsic ambiguity as given in Fig.1.
The wave functions of mental states may be now calculated on the basis of the (11-12), and they are given in the following manner

$\psi(A=+) = 0.7453$ and $\psi(A=-) = 0.6666\, e^{1.9322\, i}$

$\psi(B=+) = 0.5773 - 0.4082\, e^{1.9322\, i}$ and $\psi(B=-) = 0.4713 - 0.5269\, e^{1.9322\, i}$

The amount of doubt or of inner conflict and indetermination as induced from ambiguity of figures was calculated on the basis of the (9). It resulted

$D = 0.6667$

Since $0 \le D \le 1$, we conclude that a rather consistent value of inner conflict was induced in the examined subjects as consequence of the intrinsic ambiguity of the observed figures.
Using the (16), we calculated $D'$, the quantum interference effect, that resulted to be

$D' = 0.1112$

It resulted a rather high quantum interference effect induced from test A during resolution of test B.
In the same manner we estimated

$I_{max} = 1.9937$ ; $I_{min} = 0.0062$; $V = 0.9937$; $V^2 = 0.9874$; $D'^2 + V^2 = 0.9997$

We conclude that we examined a case of perception-cognition marked from very consistent quantum interference effects.
Finally, we must account that, according to quantum mechanics, passing from the representation of test A to test B, the subjects must realize an unitary transformation. We must inspect that this was indeed the case. In fact, on the basis of the previous results of the experimentation, we had

$$U = \begin{pmatrix} \sqrt{0.6000} & -\sqrt{0.3750} \\ \sqrt{0.4000} & \sqrt{0.6250} \end{pmatrix}$$

and

$$UU^+ = \begin{pmatrix} 0.9748 & 0.0057 \\ 0.0096 & 1.023 \end{pmatrix}$$

Therefore, also the unity was granted during such experiment.

**References**
[1] A.G. Khromov, Logical Self-Reference as a Model for Conscious Experience. Journal of Mathematical Psychology **45,** 720-731 (2001).
[2] H.P. Stapp, *Mind, Matter, and Quantum Mechanics*. (Springer Verlag, Berlin, New York,- Heidelberg 1933).
J. M. Schwartz, H.P. Stapp, M. Beauregrad, Quantum physics in neuroscience and psychology: a neurophysical model of mind-brain interaction, Phil. Trans. R. Soc.B **360**,1309-1327 (2005).
[3] J. D. Miller, The prospects for a quantum neurobiology. Neuroscience-Net, article n. 1996-011 (1996).
[4] E. Conte, O. Todarello, A. Federici, F. Vitiello, M. Lopane, A.Y. Khrennikov, *A Preliminar Evidence of Quantum Like Behaviour in Measurements of Mental States, Quantum Theory, Reconsideration of Foundations* (Vaxjio Univ. Press, 2003) pp. 679-702.
E. Conte, O. Todarello, A. Federici, F. Vitiello, M. Lopane, A.Y. Khrennikov, J.P Zbilut, Some Remarks on an Experiment Suggesting Quantum Like Behaviour of Cognitive Entities and Formulation of an Abstract Quantum Mechanical Formalism to Describe Cognitive Entity and Its Dynamics. Chaos, Solitons and Fractals **31**, 1076-1088 (2007).

[5] A. Zelinger, *On the interpretation and Philosophical Foundations of Quantum Mechanics*. (Vastokohtien todellisuus. Fetschrift for K.V. Laurakainen. U. Ketvel et al. Eds., Helsinki Univ. Press. 1966).
A. Shimony, *On Mentality, Quantum Mechanics and the Actualisation of Potentialities*. Cambridge (Univ. Press, New York 1997).
[6] N. Bohr, *The philosophical writings on Niels Bohr*. 3 vols, (Woodbridge, Conn.: OxBow Press. 1987).
[7] W. James, *Principles of Psychology*. (New York Viking Press. 1890).
[8] C.A. Meier, Ed.: *Atom and Archetype: The Pauli/Jung Letters*. (Princeton University Press. 2001) pp. 1932-1958.
[9] Y.F. Orlov, The Wave Logic of Consciousness: A Hypothesis. Int. Journ. Theor. Phys. **21**, 1 37-53 (1982).
[10] E. Conte, Exploration of Biological function by quantum mechanics, Proceedings 10[th] International Congress on Cybernetics (International Association for Cybernetics Namur-Belgique 1983) pp16-23.
[11] A.Y. Khrennikov, Linear representations of probabilistic transformations induced by cointext transitions. J. Phys. A: Math. Gen. **34**, 9965-9981 (2001).
Khrennikov, A.Y. Interference in the classical probabilistic framework. Fuzzy Sets and Systems **155**, 4-17 (2005).
A.Y. Khrennikov, Quantum-like brain: Interference of minds. BioSystems **84,** 225–241 (2006).
A.Y. Khrennikov, The principle of supplementarity: A contextual probabilistic viewpoint to complementarity, the interference of probabilities, and the incompatibility of variables in quantum mechanics. Foundations of Physics **35** (10), 1655 –1693 (2005).
[12] F. Beck, J. Eccles, Quantum aspects of brain activity and the role of consciousness. Proc. Nat. Acad. Sci. **89**, 11357-11361 (1992).
[13] E. Manoussakis, Quantum theory, consciousness and temporal perception: binocular rivalry, arxiv:0709.4516v1, 28 sep. 2007 and Found. of Physics **36** (6),795 (2006). see also
H. Atmanspacker, T. Filk, H. Romer, Quantum Zeno features of bistable perception. Biol. Cib. **90**, 33-40 (2004).
[14] G. Jaeger, A. Shimony, L. Vaidman, Two interferometric complementarities, Phys. Rev. **A51**, 54-67 (1995) .
B.G. Englert, Fringe visibility and which way information: An inequality .Phys. Rev. Lett. **81**, 5705-5709 (1998).
[15] L.A. Necker, Observations on some remarkable phenomena seen in Switzerland; and an optical phenomenon which occurs on viewing of a crystal or geometrical solid. Philosophical Magazine **3**, 329-337 (1832).

Figure 1                                                                      Figure 1

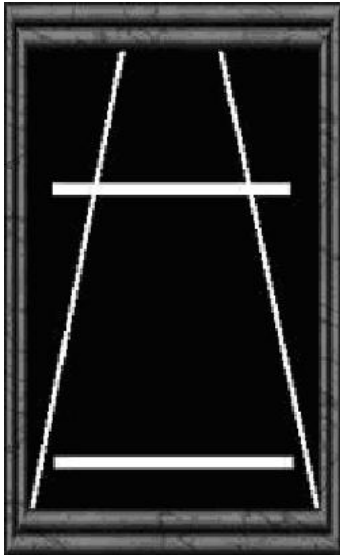

**Test B:** Are these two equal segments?

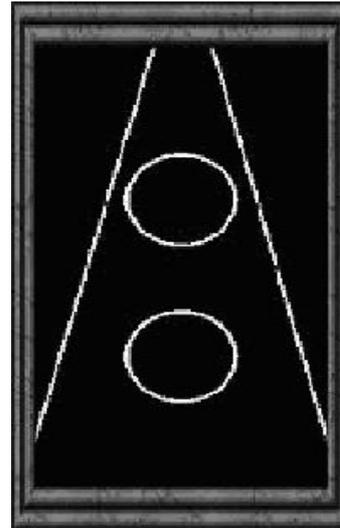

**Test A:** Are these two equal circles?

# On the Possibility That We Think in a Quantum Mechanical Manner : An Experimental Verification of Existing Quantum Interference Effects In Cognitive Anomaly of Conjunction Fallacy

Elio Conte[1,2,4], Andrei Yuri Khrennikov[3], Orlando Todarello[4], Roberta De Robertis[4], Antonio Federici[1], Joseph P. Zbilut[5]

1. Department of Pharmacology and Human Physiology – TIRES – Center for Innovative Technologies for Signal Detection and Processing, University of Bari- Italy;
2. School of Advanced International Studies for Applied Theoretical and Non Linear Methodologies of Physics, Bari, Italy;
3. International Center for Mathematical Modeling in Physics and Cognitive Sciences, MSI, University of Växjö, S-35195, Sweden;
4. Department of Neurological and Psychiatric Sciences, University of Bari, Italy;
5. Department of Molecular Biophysics and Physiology, Rush University Medical Center, 1653 W, Congress, Chicago, IL 60612, USA.

Correspondence to: Elio Conte   (email: elio.conte@fastwebnet.it) .

**Abstract**: we discuss the celebrated elaboration on Theory of Games and Economic Behaviour that was formulated by von Neumann and Morgenstern in 1944. It assumes the rationality of individual decisions in Humans. We add discussion on Tvesky and Kahneman who, in their Prospect Theory, identified instead for the first time the presence of cognitive anomalies in our mental operations. In particular the human thought and our mental operations violate rationality and does not correspond to classical, consistent and logical rules when in conjunction fallacy they violate a fundamental law of probability that a conjunction cannot be more probable than any of its constituents. It is also discussed that quantum mechanics well may represent a theory of cognitive processes, and that the existence of quantum interference effects in mental operations of the thought may explain conjunction fallacy that instead results a cognitive anomaly in the consistent framework of rules of classical probability theory. We perform an experiment of conjunction fallacy on 25 normal subjects and actually discover the existence of quantum interference effects in such cognitive dynamics. Therefore we conclude that conjunction fallacy does not represent a cognitive anomaly of our mental operations but direct expression of the fact that we think in a quantum mechanical manner.

1. Introduction

In 1944 John von Neumann and Oskar Morgenstern [1] first formulated a theory of individual decisions, the well known Theory of Games and Economic Behaviour. The basic assumption of such a theory is that the individual decisions are rational. Rationality is here intended to be the human cognition function that enables us to make use of the available information with coherence so that the human subjects use their cognitive performance to operate an optimal selection when a series of available alternatives and of fixed objectives, are given. The cognitive psychology is strongly interested to the human decisional process, and it aims to analyse the human mind and its ability to codify and to elaborate information and to solve problems. According to von Neumann and Morgenstern, cognitive psychology acknowledges that human subjects are able to perform judgements and rational choices but it takes in consideration also the importance of other cognitive components that may contribute in the final, individual choices Such factors take in consideration the perception, the formation of credences and expectations, and, generally speaking, the elaboration of mental models that shape the representation of situations that the individuals must to face up. Starting with their first papers in 1971, Tvesky and Kahneman [2] observed that the human judgement in condition of uncertainty, deviates in a systematic manner from the laws of probabilities, and, thus from rationality. In particular, a violation from such prescriptions of rationality is obtained when humans evaluate joint probability of two events to be greater than the probability of one of the events. This result places the important problem to investigate at what extent the human thoughts are rational processes. It becomes of particular relevance to understand how closely the human mental operations correspond to consistent and logical rules. The operation of conjunction (the AND operation) for judgements of likelihood represent the most fundamental passage of human rational iter. The classical fundamental law of probability fixes that a conjunction cannot be more probable than any of its components. In a great number of cases humans deviate from this rule committing what is commonly defined as conjunction fallacy. It has been confirmed in a number of experimental investigations [3]. A similar fallacy may also happen in humans in the case of the disjunction likelihood. In this case of disjunction fallacy the disjunction is judged to be less probable than any of its components. The arising problem becomes very important since it is given to investigate if there is correspondence in humans between mental operations and logical-consistent rules. Since conjunction and disjunction fallacies must be considered to be logically incorrect, it arises as conclusion that humans do not think logically. If this situation happens in so elementary mental operations (AND or OR operations) one should expect that our thinking exhibits more severe limitations when more complex thought processes are happening in our mind. Still, we know that some psychopathologies involve directly the cognitive mechanisms of the rational faculties. The inductive thought, as example, assumes great relevance in the psychopathology of the delirium that in fact is a narrative construction starting from perceptive elements and then is based on an inductive process by inference. The conjunction fallacy is a kind of inductive thought.

In conclusion, numerous studies were performed previously [4] because it is of basic interest to explore the conjunction fallacy in humans as well as in controls and in some conditions of psychopathology in order to identify some possible explanation, and, in particular, the origin of such cognitive anomaly. In this paper we will attempt a preliminary answer to this question suggesting that the source of what we call the non rational human thought process should reside in the fact that we think in a quantum mechanical manner rather than following the classical, consistent logical rules.

## 2. Quantum Mechanics and Cognition

According to the general formulation of quantum mechanics, a change in wave function happens during a measurement. It has been ascribed to an unavoidable physical interaction between measuring apparatus and the physical entity to be measured. In detail, N. Bohr in 1935 indicated that this unavoidable interaction is responsible for the uncertainty principle, and more specifically the inability to perform a simultaneous measurement of observable quantities described by non- commuting Hermitian operators. Feynman, Leighton, and Sands [5] explained that the distribution of electrons passing through a wall with two suitably arranged holes to a backstop able to detect the positions of electrons, exhibits interference. The authors explained that this interference is characteristic of wave phenomena and that the distribution of electrons at the backstop indicates that each electron acts as a wave as it passes through the wall with two holes (Fig.1). Feynman et al. explained also that if one was to introduce a procedure in order to determine through which hole the electron passes, the interference pattern is destroyed and the resulting distribution of the electrons returns to be that of classical particles passing through the two holes . Let us follow the discussion of Epstein and of Snyder. The procedure may be to introduce a strong light source behind the wall and between the two holes. It illuminates an electron as it travels through either hole (Fig.2). The general quantum interpretation of this experiment is that in determining through which hole the electron passes, the electrons are unavoidably disturbed by the photons of the light source, and this disturbance destroys the interference pattern. In quantum mechanics the act of taking a measurement affects the physical world which is being measured. However, there is still an interesting feature that is at the basis of our approach. In the same experimental arrangement one may determine which of the two holes and electron goes by using a light source that this time is placed not between the two holes  but it is placed near only one of the holes and it illuminates only the hole where it is placed (Fig.3). Also in this case in which we illuminate only one of the two holes, one determines which of the two holes the electron travels and again we have a distribution of electrons similar to that one obtains when the light source is placed between the holes. When the light source illuminates only one of the two holes, the electron passing through the other hole does not interact with photons from light source and yet interference is destroyed in the same manner as in the case in which the light source illuminates both the holes. Thus we  arrive to the conclusion that was elaborated and discussed in detail by P. Epstein in 1945 [6] and in various and fundamental  papers by D.M Snyder [7]. Quantum mechanics includes the description of some effects that cannot be ascribed to physical origin only, but they include our mental activity. Quantum mechanics is at the same time description of physical reality and Human cognition process. In fact, in the above described experiments due to Snyder, it is the knowledge factor, that is to say, our cognitive function, that plays a decisive and unavoidable role. Thus, in conclusion, quantum mechanics is at the same time science of cognition other than physical theory of matter.

D. Orlov arrived to a conclusion that, in our opinion, is very similar [8].

According to Orlov, the question arises why the logic and the language (and thus our cognitive functions) should play a so fundamental role in quantum mechanics while in classical mechanics they play only an auxiliary one. The qualitative explanation is that, though undescribed Nature certainly exists, scientific knowledge of Nature exists only in the form of logically organized descriptions. There are stages of our knowledge in which the descriptions that we realize at cognitive level become too precise that the fundamental features of cognition, that is to say of logic and of language, acquire the same importance as the features of what is being described. In other terms, at this level we cannot further separate the features of matter per se from the features of the logic and the language, and thus of cognition, used to describe it. Linked in an unavoidable role we have in quantum mechanics the physical observable $K$, with connected a set of possible numerical values $[k_1, k_2, ......]$ and a physical state $|k_i>$, and the logical-cognitive statement $\Lambda_{k_i}$ (truth of operators) :

$\Lambda_{k_i}$ : The system is in state $|k_i>$

or equivalenty $K$ is $k_i$ as true or false. To evaluate truths of statements we assign 1 to the truth value true and 0 to the truth value false. Therefore, truth operators $\Lambda_{k_i}$, with $\Lambda_{k_i}$ commuting with $K$, are projectors. Idempotents are constructed in quantum mechanics where they are, of course, called projection operators. These projection operators are used to form the basis of the propositional calculus first introduced by Birkoff and von Neumann (1936) which has since been developed into a formal structure called quantum logic (see, for example the Jauch 1968) [9]. We may also follow Eddington [10] and Bohm, Davies and Hiley, who argued that within a purely algebraic approach, which Eddington regarded as providing a structural description of physics, there are elements of existence defined, not in terms of some metaphysical concept of existence, but in the sense that existence is represented by a symbol which contains only two possibilities - existence or non-existence. In this manner the profound link between cognition and quantum mechanics still appears.

The other way to acknowledge the profound link of quantum mechanics with human cognition is that quantum theory introduces the concept of wave function and poses from its starting a net distinction between potentiality and actualisation of quantum states. On one side lies what *appears to be*, that is to say the quantum physical description of the observed actual processes while on the other side lies what is *reasoned to be* [11] that is the cognitive counterpart with potentiality marked by the mathematical expression of the wave function and its time dynamics from probabilities to actuality. Wave functions become in this manner mental entities identifying our cognitive dynamics.

Finally, let us observe that J. Eccles, F. Beck and H. Margenau [12] conducted important studies on mind nature. H. Margenau observed that quantum mechanics forces us to attribute reality to elusive entities as the probability fields, existing possibly only in our cognition. Yet, the probability fields, also being logic-cognitive construction of purely mathematical level, influence the behaviour of physical entities. Also exploring the problem at the neurological level J. Eccles was able to offer plausible arguments for mental events causing neural events analogously to the manner in which probability fields of quantum mechanics are causatively responsible for physical events. This final results still reaffirm the profound link existing between cognition and quantum mechanics mainly characterized by the abstract or mental entity of wave function.

We arrive to express this conclusion on the basis of the above discussion on the existing link between cognition and quantum mechanics.

Some mental functions as human perception and cognition can exhibit properties that result specific to quantum mechanical formalism. This conclusion opens the field to investigate quantum-probabilistic phenomena in psychology and psychophysiology. In a more general way we may say that the mathematical formalism of quantum mechanics may represent a suitable model for describing, explaining and interpreting human conscious and behavioural phenomena that pertain in particular to our thought, feeling, and so on. Of course we must be care that, introducing quantum mechanics, we change in a radical way our ontological approach to such problems. First of all we introduce the instance of non-commutativity that is at the foundation of quantum mechanics. It becomes a very incisive notion in human thought, feeling, and behaviour. In a classical vision of the world, the probabilistic character of outcomes of our observations and measurements, is due to our incomplete knowledge of the true. In the non-classical framework of quantum mechanics we enter in a new ontological perspective in which human cognition and the thought, as example, allow for an intrinsic and thus objective probabilistic character that links an intrinsic indeterminacy. It is no more matter for uncertainty due to our incomplete knowledge, in a quantum mechanical framework, such features imply an irreducible indetermination in human behaviour and our manner of thinking and feeling. In a general psychological framework we are accustomed to evaluate the human preferences and beliefs, his attitudes and the feelings. Eliciting and revealing are the usual terms by which we relate human attitudes and preferences. We admit that in humans they are well defined, determined, stable, and pre-defined in our human architecture. A very strong consequence of such classical view is that, as example, a subject is not affected during elicitation itself. This is a conclusion that, on the general plane, results to be strongly unsatisfactory in the same framework of psychologists. As example, asking to an human subject if he is angry or not, we receive a definite yes or not answer, but before posing such question to the subject, it may be neither true nor false that his state is in such definite condition. It may be in a condition of potential superposition of states as just pertains to the basic formalism of quantum mechanics.

3. **The Experimental Determination of Wave Function in Cognitive Studies**.
Let us introduce now the basic framework of our formulation.
The wave function $\psi$ of quantum mechanics represents a mental object during human cognition. We have the problem of its determination that is to say the manner in which we may acquire a posteriori a knowledge about the fact that a given system is described by a function $\psi$. Starting with 1983 [13] and 2004 [14] we considered this problem for biological and mental exploration. The basic key is that wave function $\psi$ is not an observable in the usual sense of quantum theory and, consequently, it cannot be measured in the usual sense of this word. On the other hand, it may be determined provided that one has an ensemble of similarly prepared systems, each describable by the same function $\psi$. To this purpose [13, see also 14] one measures an arbitrary but complete set of observables that we call A on a large number of systems and get a statistical distribution of eigenvalues $A_n$. This approach determines the absolute values $|a_n|$ of the coefficients of the decomposed wave function

$$\psi = \Sigma_n a_n \psi_{A_n} \quad \text{with} \quad \Sigma_n |a_n|^2 = 1 \tag{1}$$

In order to determine $\psi$ one must know not only the absolute values of the coefficients but also the
phases

$$a_n = |a_n| e^{i\alpha_n} \tag{2}$$

Therefore, we have to repeat the same procedure again by measuring another complete set of observables B on a large number of systems to obtain an independent decomposition

$$\psi = \Sigma_m b_m \psi_{B_m} \quad \text{with} \quad \Sigma_m |b_m|^2 = 1 \tag{3}$$

and determine $|b_m|$. Let us admit that we find a number $N$ and $N'$ of coefficients different from zero, respectively. In the case $N > N'$ we introduce the decomposition

$$\psi_{B_m} = \Sigma_n c_{mn} \psi_{A_n} \tag{4}$$

in the (3), and comparing it with the (1), we have that

$$a_n = \Sigma_m b_m c_{mn} \tag{5}$$

We have a set of $N$ complex equations, $2N$ real equations for $N + N'$ phases to be determined. If $N' > N$, we have to interchange $A$ with $B$. If they result not soluble, it means that we have not a pure quantum state but a mixture. If instead they result dependent so that the phases cannot be determined uniquely, this means that A and B are not independent of each other and consequently we have to make a new choice for observables.

In conclusion, we have delineated the method to realize a determination of wave functions of mental states starting directly from experimental data. The approach must be clear. It is possible to determine the wave function only post factum and considering an ensemble of similarly prepared subjects. In any case the concept of measurement of observables applied to single systems cannot be confused with the estimation of a wave function. This last is expected to represent the system but remains unknown. In other terms, the function $\psi$ is not observable in itself and thus cannot be measured. However, being subject to probabilistic interpretation, it may be estimated statistically.
This is to say that we make only an a posteriori statistical reconstruction of it.

**4. The Quantum Formulation of Mind Entities during Human cognition**
According to Eccles, Beck, and Margenau [12] the mind is a field of probability. In quantum mechanics, the abstract probability fields causally preside over the advent of events in nature. Mathematically, a quantum state $\psi$ (wave function) is a vector in Hilbert space $H$ over some field. Such a state has a probabilistic content, the vector that represents it, has to be of length one, that is to say
$<\psi|\psi>^2 = 1$ (where $<\cdot/\cdot>$ denotes the inner product in $H$). The basic key of intrinsic indetermination in quantum mechanics is the principle of superposition. It states that for any two
states $\varphi_1$ and $\varphi_2$ of $H$, there exist at least one other state $\psi$ of $H$ that can be written as a linear combination of the first two :
$$\psi = c_1\varphi_1 + c_2\varphi_2 \qquad (6)$$
with
$$|c_1|^2 + |c_2|^2 = 1 \; ; \; <\varphi_i|\varphi_j> = \delta_{ij} \; (i = 1,2 \; ; j = 1,2) \qquad (7)$$
and $|c_1|^2$ represents probability for state $\varphi_1$, $|c_2|^2$ represents probability for state $\varphi_2$.
Mind states are represented by the quantum superposition principle during cognition.
In our quantum model of mind entities (for details see also [15]) we consider that an human subject can potentially have multiple mind representations of a given situation, also if actually he can attend to only one representation at any given time. In this quantum mechanical framework we distinguish a potential and an actual or manifest state of consciousness during cognition. The state of the potential consciousness will be represented by a vector in Hilbert space. If we indicate for example a bidimensional case of a cognitive decision situation with potential states $|+>$ and $|->$, in relation to a dichotomous observable $A = +,-$, the potential state of cognition awareness will be given by the superposition
$$\psi = a|+> + b|-> \qquad (8)$$
Here, $a$ and $b$ represent probability amplitudes so that $|a|^2$ will give the probability that the state of cognitive awareness, represented by $|+>$, will be finally actualised, that is to say cognitively manifested during decision. Conversely $|b|^2$ will represent the probability that state $|->$ of cognitive awareness will be actualised, cognitively manifested during decision. It will be $|a|^2 + |b^2| = 1$.
The potential state, given in (8), represents in some manner the condition of doubt or of inner conflict or of intrinsic indetermination of the human subject during cognition and in relation to the posed question ( $A = +,-$ ).
The amount of doubt of the subject (see also [8]) or his inner conflict or indetermination, in relation

to the posed question ($A = +,-$), is given by

$$D = 1 - \left| |a|^2 - |b|^2 \right| ; \quad 0 \leq D \leq 1 \tag{9}$$

When an actual or manifest state of cognition is realized during decision, the (8) is reduced to $|+>$ with probability $|a|^2$ or to $|->$ with probability $|b|^2$.

As neurophysiological counterpart of the present quantum mechanical model of mind entities, as also previously outlined in [16], we admit that, when a conscious decision or observation happens, the actual event that in correspondence is realized in consciousness, is linked to a particular neural correlate brain state. In this manner, in the (8), $|+>$ and $|->$ represent two possible states having two distinct neural correlates.

For brevity we will not consider here the case of the evolution in time of the state of potential consciousness, see the [15] for details.

A final comment may be now added to the previous formulation.

The principle of superposition of states, discussed ontologically in the previous sections, implies that the state space is non- Boolean. Therefore, mind states during cognition pertain to a non-Boolean state space in this model. The arising conclusion is that at least some perceptive- cognitive systems have such quantum abilities. The brain should result to emulate quantum dynamics at least under some conditions. Such an emulation would allow for a three-valued logic in human cognition: true, false and the potential superposition of true and false. This could relate the peculiar human cognitive ability to hold contradictory notions in mind simultaneously, although usually there is collapse to one state or the other. This ability to see things from "opposite" views might have been valuable in the development of sociability, empathy and even cognitive innovation which seems to depend on seeing things in a radically different way as compared to social or cultural norms. The final important feature of the present cognitive model ,relates the nature of mind entities. If the previous model is confirmed, we have to conclude that they, at least under our conditions of experimentation, operate by quantum probabilities and analysing (even unconsciously) probabilities of various alternatives. They work directly with mental wave functions and probabilistic amplitudes.

**4.Theoretical Description of the Experiment on Conjunction Fallacy.**

According to quantum mechanical results that we have given in the previous section, it becomes of interest to establish the dynamics of such cognitive mechanism when two or more questions $A$, $B$,... are posed to an human subject. As previously outlined, we associate to every cognition situation a relating cognitive observable that we denote by $A$, $B$,...that we consider to act on $H$. We may study more that one cognitive decision, say ($A = +,-$) and ($B = +,-$) to be posed to the subject. We know that a key question in quantum mechanics is whether the corresponding observables are or not commuting operators in $H$, i.e., whether $AB = BA$ or not.. From a formal view point we have discussed the case $AB \neq BA$ by the formulas (1-5) of the previous sections. From a cognitive view point we may expect that if $A$ and $B$ commute, a decision $A$ will not affect the subsequent decision on $B$. The situation is completely different in the case in which observables $A$ and $B$ do not commute. This is the most interesting case in our psychology studies. In fact, in this case the predictions of quantum mechanics differ radically from those of the classical probabilistic model. The reason is that in this case the quantum probability calculus generates cross- terms also called the interference terms. In our formulation these cross- terms are the signature of an existing intrinsic indetermination, of an intrinsic doubt , of an inner conflict that characterizes the cognitive status of the subject in the situation of $A$ and B posed questions. Generally speaking, admitting that questions $A$ and $B$ have the same number $n$ of possible choices, considering again the (1-5), we obtain

$$p_B(b_n) = \left(\sum_{j=1}^{n} a_j c_{nj}\right)^2 = \sum_{j=1}^{n} a_j^2 c_{nj}^2 + 2\sum_{j \neq j'} \left[(a_j c_{nj})(a_{j'} c_{nj'})\right] \tag{10}$$

where the first term in the (10) indicates the classical term of Bayes probability to which it is added the quantum interference term that is the quantum expression of an irreducible intrinsic indetermination during cognition of the subject. In brief, the (10) represents a very powerful instrument to test if our cognitive apparatus follows or not quantum mechanics. In fact, if quantum mechanics has a role in the investigated cognitive process, we have a violation of the classical Bayes formula for conditional probability and we may quantify the quantum interference term that otherwise results to be zero.. Let us examine the case of two dichotomous observables and thus involving two decisions, ($A = +,-$) and (B=+,-). We have the following wave functions

$$\psi(A=+) = \sqrt{P(A=+)}\, e^{i\varphi_1}, \qquad \psi(A=-) = \sqrt{P(A=-)}\, e^{i\varphi_2},$$
$$\psi(B=+) = \sqrt{P(B=+)}\, e^{i\vartheta_1}, \qquad \psi(B=-) = \sqrt{P(B=-)}\, e^{i\vartheta_2} \tag{11}$$

According to the (1-5) we have that

$$\psi(B=+) = \sqrt{P(B=+/A=+)}\, \psi(A=+) - \sqrt{P(B=+/A=-)}\, \psi(A=-)$$
$$\psi(B=-) = \sqrt{P(B=-/A=+)}\, \psi(A=+) + \sqrt{P(B=-/A=-)}\, \psi(A=-) \tag{12}$$

The square modulus of such probability amplitudes gives

$$P(B=+) = P(A=+)P(B=+/A=+) + P(A=-)P(B=+/A=-)$$
$$- 2\sqrt{P(A=+)P(A=-)P(B=+/A=-)P(B=+/A=+)}\, \cos(\varphi_1 - \varphi_2) \tag{13}$$

and

$$P(B=-) = P(A=+)P(B=-/A=+) + P(A=-)P(B=-/A=-)$$
$$+ 2\sqrt{P(A=+)P(A=-)P(B=-/A=+)P(B=-/A=-)}\, \cos(\varphi_1 - \varphi_2) \tag{14}$$

We see that in the (13) and (14), in addition to the classical Bayes formula for conditional probabilities, a quantum interference term appears that acclaims the presence and the role of quantum effects at the cognitive level of the subject. The experimental determination of probabilities
$P(B = +)$, $P(B = -)$, $P(A = +)$, $P(A = -)$, $P(B = + / A = +)$, $P(B = + / A = -)$
enables us to calculate the interference term

$$\cos \omega = \frac{P(B=+) - P(A=+)P(B=+/A=+) - P(A=-)P(B=+/A=-)}{2\sqrt{p(B=+)p(A=+/B=+)p(B=-)p(A=+/B=-)}} =$$
$$= \frac{\Delta p}{2\sqrt{P(A=+)P(B=+/A=+)P(A=-)P(B=+/A=-)}} \tag{15}$$

and the phase $\omega = \varphi_1 - \varphi_2$ by which the a posteriori determination of the quantum wave function
(see the 12) is realized.
We have to add some specifications:
1) Let us reconsider the text that we wrote after the (5) in section 3 of the present paper. If we arrive to the unique determination of phase $\omega$, we are sure that we are not in presence of A and B dependent observables. Otherwise, A and B are not independent of each other and consequently we have to make a new choice for observables in order to ascertain the presence of an actual quantum interference effect in our experiment.
2) Let us consider how the (13) and the (14) change in a radical way our manner to intend a cognitive process in psychology and the human rationality. In fact, if there is correspondence between Human's mental operations and classical logical-consistent rules, we have always that

as example
$$P(B=+) > P(A=+)P(B=+/A=+) \qquad (16)$$
In this manner we conclude that human thoughts are not rational in the cases in which mental operations violate such consistent and logical classical rule. Instead , if we accept that we think in a quantum mechanical manner, it is the (13) to have its validity and the presence in (13) of the quantity $\cos\omega$, that may be positive or negative [see also previous conclusions on this matter in 4], enables the violation of (16) without invoking no more Human non rationality. In this case we have not to speak of cognitive anomaly in conjunction fallacy but about the net behaviour of human cognitive apparatus to think in a quantum mechanical manner rather in a classical one as constantly admitted in psychological cognitive studies. Obviously, the verification of such a result opens important perspectives in the field of psychology and psychopathology, in neurology, in philosophy of mind and in general in all studies relating decision theory, human behaviour and still social sciences. Ascertaining the presence of quantum interference in conjunction fallacy means that we have to reformulate at a cognitive level the problem of what we must intend for human rationality starting from the initial point that we could think in a quantum mechanical manner and, in fact, this thesis has profound implications at very different fields of studies.

**5. The Performed Experiment on Conjunction Fallacy.**
The reason to perform an experiment on conjunction fallacy was largely explained in the previous section. In our laboratories we investigated a group of 25 graduate students in psychology with age included between 22 and 27 years old, and near equal distribution of females and men. All the subjects were previously submitted to our routine controls in order to be accepted as normal subjects for our experimentation.
The posed sentences were the following:
Question A : Within November 2009 the consumption of cigarettes will decrease about 15% among the young men in our country. Yes, A=+ ; Not A=- .
Question B and A :
The price of the cigarettes will increase of about 1 euro (B=+,yes ; B=-,Not) and within November 2009 the consumption of cigarettes will decrease about 15% among the young men in our country (A=+,yes; A=-,Not).
To each subject was asked to select the event judged more probable between Question A or Question B and A, and indicating in the second case his choice as B=+, A=+ or B=+,A=-, or B=-, A=+, B=-, A=-. Not that we have interchanged A with B respect to the theoretical exposition of the previous section, but it does not change the foundation of our exposition.
The results of the experiment were as it follows:
On a total of 25 subjects , 8 subjects judged more probable the event consisting of the only question A while instead 17 subjects selected the question B and A. It is thus confirmed the tendency of subjects to retain more probable the conjunction that any of its constituents (conjunction fallacy).

P(question A) = $\frac{8}{25}$ = 0.320 ; P(question B and A)= $\frac{17}{25}$ = 0.680 with $\chi^2$ = 25.90 (99.9%)

For P(A=+) and P(A=-) we had:
P(A=+)= $\frac{1}{8}$ = 0.125 and P(A=-)= $\frac{7}{8}$ = 0.875 ;
In addition we had :
P(B=+)= $\frac{13}{17}$ = 0.765 ; P(B=-)= $\frac{4}{17}$ = 0.235 ; P(A=+/B=+)= $\frac{5}{13}$ = 0.385;

P(A=-/B=+)= $\frac{8}{13}$ = 0.615; P(A=+/B=-)= $\frac{2}{4}$ = 0.500; P(A=-/B=-)= $\frac{2}{4}$ = 0.500.

In conclusion it resulted :
P(A=+)=0.125   and P(B=+)P(A=+/B=+)=0.294 . We had
P(A=+)<P(B=+)P(A=+/B=+)  -   Conjunction Fallacy with $\chi^2 = 7.71$ (99.4%)
Calculation of the Quantum Interference Effect. For the case of P(A=+) , on the basis of the (15) we had
$\cos\omega$ =-0.78623   and   $\chi^2 = 19.88$ (99.9%) for $\Delta p$.
For the case of P(A=-) we had
$\cos\omega = 0.60933$
In this manner we had confirmation of existing quantum interference effects in the investigated cognitive process, Therefore, it follows the conclusion that conjunction fallacy does not represent an Human cognitive anomaly. It reflects the fact that we reason in a quantum mechanical manner.

**References**
[1] John von Neumann and Oskar Morgenstern, Theory of Games and Economic Behaviour, Princeton University Press; 3 edition (May 1, 1980);
[2] D. Kahneman, A. Tversky, Choice, Values and Frames, Cambridge University Press, 2000;
[3] S. Stolarz-Fantino, E. Fantino, D.J. Zizzo, The conjunction fallacy: New evidence for robustness, American Journal of Psychology, 116(1),15-34,2003;
[4] F.J Costello . A unified Account of conjunction fallacy and disjunction fallacies in people's judgments of likelihhod, www.cogsci.rpi.edu/CSJarchive/Proceedings/2005/docs/p494.pdf, and references therein;
For a quantum discussion see also
R. Franco, Quantum mechanics and rational ignorance, arXiv.0702163 –gen.phys;
[5] R. Feynman, R.B. Leighton, M. Sands, The Feynman lectures on physics: quantum mechanics, Reading Massachusettts:Addison –Wesley, 1965;
[6] P.S. Epstein, The reality problem in quantum mechanics, American Journal of physics, 13, 3, 127-136, 1945;
[7] D.M. Snyder, On the nature of the change in the wave function in a measurement in quantum mechanics arxiv : quant-ph/9601006; On the quantum mechanical wave function as a link between cognition and the physical world: a role for psychology,  Cogprints, ID code 2196, 30 Apr.2002, last modifies 12 Sept.2007
[8] V. Orlov, Peculiarities of Quantum Mechanics : origins and meaning, arxiv: quant-ph /9607017;
Y.F. Orlov, The Wave Logic of Consciousness: A Hypothesis. Int. Journ. Theor. Phys. 21, 1 37-153, 1982;
[9] J.M. Jauch,  Foundations of Quantum Mechanics, Addison Wesley, Massachusetts1968;
[10] A.S.  Eddington, Fundamental Theory, Cambridge 1946; see also D.J. Bohm, P.G. Davies, B.J. Hiley, Algebraic Quantum Mechanics and Pregeometry, arXiv: quant-ph/0612002;
[11] M. Epperson, Bridging necessity and contingency in quantum mechanics: The scientific rehabilitation of process metaphysics, ctr4process.org/publications/.../LSI05/Epperson_CPS Paper_REVISED.pdf
[12] F. Beck, J. Eccles, Quantum aspects of brain activity and the role of consciousness, Proc. Natl. Acad. Sci. USA, 89, 11357-11361,1992;
H. Margenau, The miracle of existence, Armando Editore, 1984;
[13] E.Conte, Exploration of Biological function by quantum mechanics, Proceedings 10[th] International Congress on Cybernetics (International Association for Cybernetics Namur-Belgique
1983) pp16-23.

[14] A.Y. Khrennikov, Linear representations of probabilistic transformations induced by cointext
transitions. J. Phys. A: Math. Gen. 34, 9965-9981, 2001;
Khrennikov, A.Y. Interference in the classical probabilistic framework. Fuzzy Sets and Systems 155, 4-17, 2005;
A.Y. Khrennikov, Quantum-like brain: Interference of minds. BioSystems 84, 225–241, 2006;
A.Y. Khrennikov, The principle of supplementarity: A contextual probabilistic viewpoint to complementarity, the interference of probabilities, and the incompatibility of variables in quantum
mechanics. Foundations of Physics 35 (10), 1655 –1693, 2005;
[15] Elio Conte, Andrei Yuri Khrennikov, Orlando Todarello, Antonio Federici, On the Existence of Quantum Wave Function and Quantum Interference Effects in Mental States: An Experimental Confirmation during Perception and Cognition in Humans   arXiv:0807.4547 gen-ph;
Elio Conte, Orlando Todarello, Antonio Federici, Francesco Vitiello, Michele Lopane, Andrei Khrennikov, Joseph P. Zbilut, Some Remarks on an Experiment Suggesting Quantum Like Behavior of Cognitive Entities and Formulation of an Abstract Quantum Mechanical Formalism to Describe Cognitive Entity and its Dynamics, Chaos, Solitons and Fractals, 31, n. 5, 1076-1088, 2006
[16] E. Manoussakis, Quantum theory, consciousness and temporal perception: binocular rivalry, arxiv:0709.4516v1, 28 sep. 2007 and Found. of Physics 36 (6), 795, 2006; see also
H. Atmanspacker, T. Filk, H. Romer, Quantum Zeno features of bistable perception. Biol. Cib. 90,
33-40, 2004.

**Figures**

**Fig. 1**

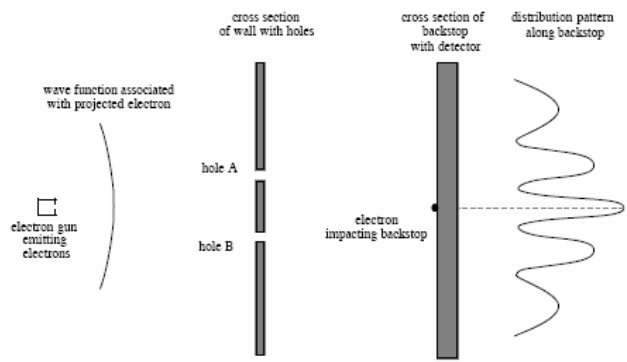

Figure 1
Two-hole gedankenexperiment in which the distribution of electrons reflects interference in the wave functions of electrons.
(Gedankenexperiment 1)

**Fig. 2**

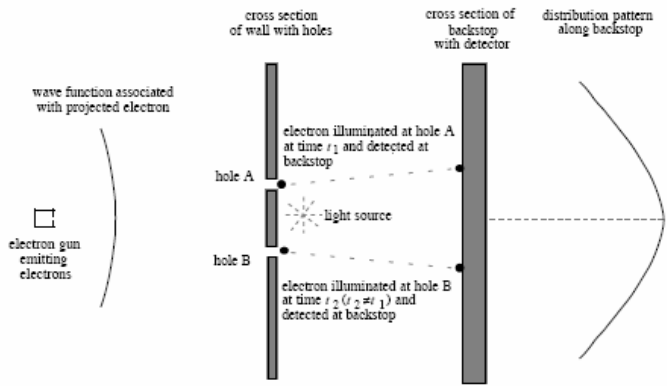

Figure 2
Two-hole gedankenexperiment with strong light source.
(Gedankenexperiment 2)

**Fig. 3**

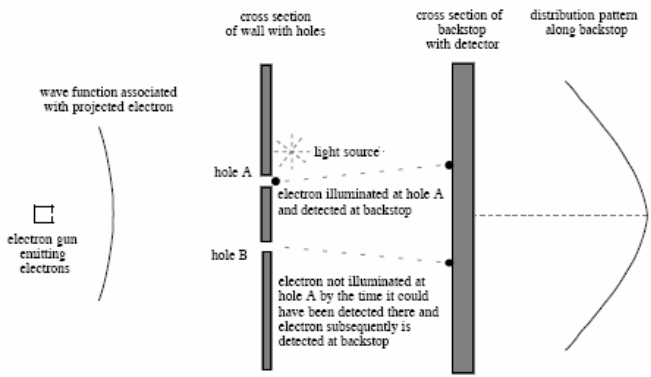

Figure 4
Two-hole gedankenexperiment with strong light source illuminating only one hole.
(Gedankenexperiment 3)

The above Figures 1,2,3 are taken from D.M. Snyder paper quoted in [7], arxiv : quant-ph/9601006.